\documentclass[12pt,a4paper]{article}
\usepackage{amssymb,amsmath,graphicx,subcaption,cite}

\usepackage{epstopdf}
\usepackage[utf8]{inputenc}
\usepackage[english]{babel}

\begin{document}

	
\begin{center}
   {\Large\bf Nonequilibrium Multiple Transitions in the Core-shell Ising Nanoparticles Driven by Randomly Varying Magnetic Fields }
\end{center}
\vskip 1 cm
\begin{center} 
    Erol Vatansever$^{1,*}$ and Muktish Acharyya$^{2, \dag}$
   
   \textit{$^1$Department of Physics, Dokuz Eylul University, TR-35160, Izmir-Turkey}\\
   \textit{$^2$Department of Physics, Presidency University, 86/1 College Street, 
           Kolkata-700073, India}\\
\vskip 0.2 cm
   {E-mail$^*$:erol.vatansever@deu.edu.tr}\\
   {E-mail$^\dag$:muktish.physics@presiuniv.ac.in}
\end{center}
\vspace {1.0 cm}
	
\noindent {\bf Abstract:} The nonequilibrium behaviour of a core-shell nanoparticle has been studied by Monte-
Carlo simulation. The core consists of Ising spins of $\sigma=1/2$ and the
shell contains Ising spins of $S=1$. The interactions within the core and in the shell are considered
ferromagnetic but the interfacial interaction between core and shell is antiferromagnetic. 
The nanoparticle system is kept in open boundary conditions and is driven by randomly varying (in time but uniform over the space) magnetic field. 
Depending on the width of the randomly varying field and the temperature of the system,
the core, shell and total magnetization varies in such a manner that the time averages
vanish for higher magnitude of the width of random field, exhibiting a dynamical
symmetry breaking transitions. The susceptibilities get peaked at two different
temperatures indicating nonequilibrium multiple transitions. The phase boundaries of the 
nonequilibrium multiple transitions are drawn in the plane formed by the axes of temperature and the width
of the randomly varying field. Furthermore, the effects of the core and shell thicknesses on the multiple 
transitions have been discussed.

\vskip 4cm

	
\newpage
\section{Introduction}\label{introduction}
\vspace {0.5 cm}

The random field Ising model (RFIM) including quenched random magnetic field has attracted a considerable interest in the 
last four decades \cite{Imry, Ahorony}. Despite of its simplicity, many problems in statistical physics and condensed matter physics can be studied
by means of RFIM.  Experimental examples include diluted antiferromagnets $\mathrm{Fe_xZn_{1-x}F_2}$ \cite{Belanger, King}, 
$\mathrm{Rb_2Co_xMg_{1-x}F_4}$ \cite{Ferreira, Yoshizawa}, $\mathrm{Co_xZn_{1-x}F_2}$ \cite{Yoshizawa} in a magnetic field and 
colloid-polymer mixtures \cite{Vink, Annunziata}. From the theoretical point of view, thermal and magnetic phase transition 
properties of the static RFIM have been investigated by a wide variety of techniques such as Molecular Field Theory (MFT) \cite{Ahorony, Schneider, Andelman, Mattis, Kaufman, Queiros}, Effective-Field Theory (EFT) \cite{Borges, Sarmento, Sebastianes, Kaneyoshi, Albuquerque, Akinci, Karakoyun} and Monte Carlo (MC) simulations \cite{Landau, Machta, Fytas1, Fytas2, Gofman, Fytas3, Fytas4, Fytas5}.  These theoretical studies show that 
different random-field distributions may lead to different physical outcomes, and thereby the existence of a 
quenched impurity  in magnetic field  has an important role in material science. They also indicate that our understanding 
of equilibrium critical phenomena associated with the RFIM has reached a point in which the satisfactory results are available. The readers 
may refer to \cite{Fytas6} for a detailed review of recent developments in the RFIM. 
However, far less is known for the physical mechanisms underlying 
the out of equilibrium phase transitions of the Ising systems in the presence of a randomly varying magnetic field.

Stationary state properties of a randomly driven Ising ferromagnet has been investigated by benefiting from the Glauber 
dynamics \cite{Hausmann}. Based on the MFT calculations, it is found that the system shows a first-order phase transition 
related to dynamic freezing. Paula and Figueiredo \cite{Paula} have attempted to study dynamical behavior of the Ising model in a quenched random magnetic field, with a bimodal distribution for the random fields. The dynamics of the system has been defined in terms of 
Glauber type stochastic process. It is obtained that the magnetic field values leading to first-order transitions are greater 
than the corresponding fields at equilibrium. One of us \cite{Acharyya} has focused on the two dimensional Ising model in the presence of randomly varying magnetic field to understand how the randomly changing magnetic affects the physical 
properties of the system. By benefiting from both MFT and MC simulations, it is reported that the time-averaged magnetization 
disappears from a nonzero value depending upon the values of the
 width of randomly varying field and the temperature. Nonequilibrium 
phase transition properties in a three-dimensional lattice system with random-flip kinetics have been elucidated in detail 
by using MC simulations \cite{Crokidakis}. One of the remarkable findings is that the system displays a first-order phase 
transitions located at low temperature region and large disorder strengths, denoting a nonequilibrium
tri-critical point at a finite temperature. Moreover, nonequilibrium phase transitions and stationary-state solutions of a three-dimensional
random-field Ising model under a time-dependent periodic external field have been investigated within the framework of EFT with single-site 
correlations \cite{Yuksel}. The amplitude of the external is chosen such that they will be according to bimodal and 
trimodal distribution functions. It is shown that the system explicits unusual and interesting behaviors depending on type of the 
magnetic field source. The readers may refer to \cite{Acharyya_Chakrabarti} for a review of the dynamic phase 
transitions and hysteresis phenomena observed in different kinds of magnetic systems. To the best our knowledge, all of the studies mentioned above have been dedicated to bulk materials in the presence of randomly varying magnetic field. There are, however, only a few studies  regarding the random magnetic field effects on the core/shell 
nanoparticle systems including surface and finite size effects \cite{Zaim, Zaim2, Kaneyoshi2}.  It is a well known fact that when 
the physical size of an interacting magnetic system reduces to a characteristic length, surface effects begin to show themselves 
on the system, and as a result of this, some interesting behaviours can be  observed, differing from those of bulk 
systems \cite{Berkowitz}. Two notable examples are $\mathrm{Co/CoO}$ \cite{Meiklejohn_1, Meiklejohn_2} and $\mathrm{Mn/Mn_3O_4}$ \cite{Alvarez}, where the physical properties of the nanoparticles sensitively depend on its own chemical characters. In this paper, we will consider the core/shell ferrimagnetic nanoparticle system driven by a randomly 
varying magnetic field. More specifically, our motivation is to understand how randomly changing field affects the thermal and magnetic properties of a nanoparticle, by means of an extensive MC simulation. In a nutshell, our simulation results indicate that the present system exhibits multiple dynamic phase transitions depending on the chosen values of the external magnetic field width and the temperature of the system.

The rest of the paper is organized as follows: In Sec. (\ref{Model}), we present the model and simulation details. The numerical 
findings and discussion are  given in Sec. (\ref{Results}), finally Sec. (\ref{Conclusion}) is dedicated to a 
brief summary of our conclusion.

\section{Model and Simulation Details}\label{Model}
We consider a cubic with thickness $(L=L_c+L_{sh})$ ferrimagnetic nanoparticle composed of a spin-1/2 ferromagnetic 
core which is surrounded by a spin-1 ferromagnetic shell layer. Here, $L, L_c$ and $L_{sh}$ are the total, core and shell 
thicknesses of the particle, respectively.  At the interface, we 
define an antiferromagnetic interaction between core and shell spins. The nanoparticle 
is exposed to a time dependent randomly varying magnetic field. The Hamiltonian describing 
our system can be written as follows:

\begin{equation}\label{Eq1}
 H=-J_{c}\sum_{\langle ij\rangle} \sigma_{i}\sigma_{j}-J_{sh}\sum_{\langle kl\rangle} 
 S_{k}S_{l}-J_{int}\sum_{\langle ik\rangle} \sigma_{i}S_{k}-h(t)\left(\sum_{i}\sigma_{i}+
 \sum_{k}S_{k}\right)
\end{equation}
here $\sigma_{i}=\pm 1/2$ and $S_{k}=\pm 1,0$ are spin variables corresponding to the 
core and shell parts of the particle, respectively. $J_c$  and $J_{sh}$ denotes the 
ferromagnetic spin-spin interactions in the core and shell components of the system 
while $J_{int}$ is the antiferromagnetic interaction at the interface of 
the particle. The symbol  $\langle \cdots \rangle$ represents the nearest neighbor 
interactions in the system. $h(t)$ is the randomly varying magnetic field 
(in time but uniform in space). The time variation of $h(t)$ can be given 
as follows \cite{Acharyya}:

\begin{equation}\label{Eq2}
 h(t)=
\begin{cases}
   h_0r(t)        & \text{for } t_0<t<t_0+\tau \\
   0        & \text{otherwise }
  \end{cases}
\end{equation}
where $r(t)$ is the random number which is distributed uniformly between -1/2 and 1/2. 
Thereby, the field $h(t)$ varies randomly from $-h_0/2$ to $h_0/2$ and,
\begin{equation}\label{Eq3}
 \frac{1}{\tau}\int_{t_0}^{t_0+\tau}h(t)dt=0
\end{equation}

For the sake of simplicity, $J_{sh}$ is fixed to unity throughout the 
simulations, and the remaining system parameters are normalized with $J_{sh}$. In 
order to study thermal and magnetic properties of the system, we employ Monte Carlo 
simulation with single-site update Metropolis algorithm \cite{Newman, Binder} on a $L \times L \times L$ simple cubic lattice. We apply boundary conditions such that they are free in all directions of the particle. 
We note that such type of a boundary condition is an
appropriate choice for considered finite small system. Let us briefly summarize the simulation protocol we follow here: The system is in contact with an isothermal heat bath at a reduced temperature $k_BT/J_{sh}$, where $k_B$ is the Boltzmann constant. Spin configurations were produced by selecting the spins randomly through the lattice, and the single-site update Metropolis algorithm was used for each considered spin. This process was repeated $L^3$ times, which  also defines a MC step per site.

Using the above scheme we simulated the nanoparticle system. 100 independent 
initial configurations have been generated to get a satisfactory statistics. For each initial spin configuration, the 
first $5\times 10^5$  MC steps have been discarded for thermalization process, and the numerical data were measured during the following $5\times 10^5$ MC steps. Based on our detailed test calculations, it is possible to say that this number of transient steps is found to be enough for thermalization of the particle. We have verified that higher values of transient steps does not change the 
outcomes reported here. Error bars have been obtained using the jackknife method \cite{Newman}. The main quantity of interest is the time-averaged magnetization, which is defined as follows:

\begin{equation}\label{Eq4}
 Q_{\alpha}=\frac{1}{\tau}\int_{t_0}^{t_0+\tau} m_{\alpha}(t)dt
\end{equation}
 
\noindent where $\alpha = c, sh$ and $T$ corresponding to the core and shell components of the 
particle and the overall of the system. $m_{\alpha}(t)$ is the time-dependent magnetization, which can be given as follows:
\begin{equation}\label{Eq5}
 m_{c}(t)=\frac{1}{N_{c}}\sum_{i}^{N_{c}} \sigma_i, \quad m_{sh}(t)=\frac{1}{N_{sh}}\sum_{i}^{N_{sh}} S_i, \quad  
m_{T}(t)=\frac{N_{c}m_{c}(t)+N_{sh}m_{sh}(t)}{N_{c}+N_{sh}}
 \end{equation}

\noindent here $N_{c}=L_c^3$ and $N_{sh}=L^3-L_c^3$ denotes the total number of spins lying in the core and shell parts of the system, respectively. We select the number of core and shell spins as $N_c=10^3$ and $N_{sh}=14^3-10^3$, such that it allows us to create a core-shell nanocubic particle
with shell thickness $L_{sh}=2$ unless otherwise stated. We also define two additional order parameters for the core and shell layers of the particle as follows:  
\begin{equation}\label{Eq6}
O_c=\frac{N_c}{N_c+N_{sh}}Q_c, \quad \quad \quad O_{sh}=\frac{N_c}{N_c+N_{sh}}Q_{sh}.
 \end{equation}

\noindent {By benefiting from the definitions given in Eq. (\ref{Eq6}), the total order parameter can be defined by: $O_{T}=O_{c}+O_{sh}$. To estimate the pseudo-critical transition temperature for a finite-size system as a function of the external field, it is useful to focus on  
the scaled variances of the dynamic order parameters \cite{Robb},
\begin{equation} \label{Eq7}
 \chi_{\alpha}=N_\alpha\left(\langle O_\alpha^2 \rangle- \langle |O_\alpha|\rangle^2 \right),
\end{equation}
where $N_{T}=N_c+N_{sh}$ is the total number of spins in the system. In addition to the 
Eq. (\ref{Eq7}), we also measure the scaled variance of the total energy (which can be considered as heat-capacity in equilibrium system) of the particle including the 
cooperative part as follows:
\begin{equation}\label{Eq8}
 \chi_E=N_{T}\left(\langle E^2 \rangle- \langle E\rangle^2 \right),
\end{equation}
where $E$ is the time-averaged energy of the particle per site, which is defined as follows:
\begin{equation}\label{Eq9}
E=-\frac{1}{\tau N_{T}}\int_{t_0}^{t_0+\tau} \left[J_{c}\sum_{\langle ij\rangle} \sigma_{i}\sigma_{j}+J_{sh}\sum_{\langle kl\rangle} 
 S_{k}S_{l}+J_{int}\sum_{\langle ik\rangle} \sigma_{i}S_{k}\right]dt.
\end{equation}

\section{Results and discussion}\label{Results}

 Figure \ref{Fig1} shows the time series of core $m_{c}(t)$, shell $m_{sh}(t)$ and total $m_{T}(t)$ magnetizations
at a considered value of temperature and for two different values of randomly field width $h/J_{sh}$. All numerical 
results given here are obtained for $k_{B}T/J_{sh}=2.0$. It is clear from the Figure \ref{Fig1}(a) that when the 
external field width is relatively small (for example $h/J_{sh}=1.0$) time dependent magnetizations can not give a response to changing magnetic 
field simultaneously. Thereby, the system remains in a dynamically symmetry broken phase where 
the instantaneous magnetizations oscillate around a non-zero value. If $h/J_{sh}$ gets 
bigger, time dependent magnetizations tend to give an answer to randomly changing magnetic field. Therefore, the system 
remains in a dynamically symmetric phase where the magnetizations oscillate around zero value, as depicted in Figure \ref{Fig1}(b). The results aforementioned here indicate that dynamics of the core/shell nanoparticle sensitively depends on the chosen external field width value. It should be noted that such type of an observation has been reported for the bulk spin-1/2 Ising models 
driven by a randomly changing magnetic field in both two and three-dimensional space \cite{Acharyya, Yuksel}.

\begin{figure}[htbp]
\includegraphics*[width=7.0 cm]{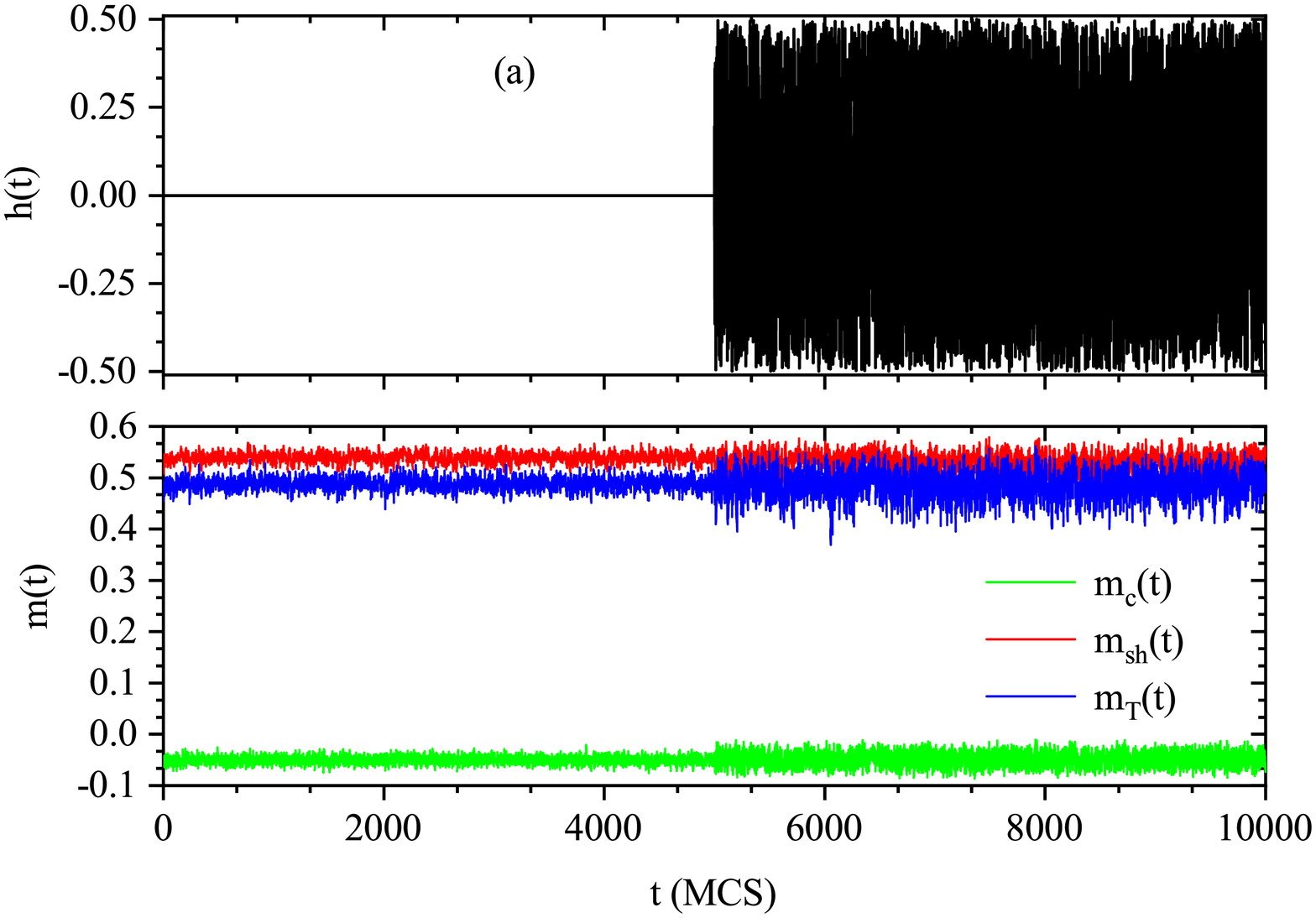}
\includegraphics*[width=7.0 cm]{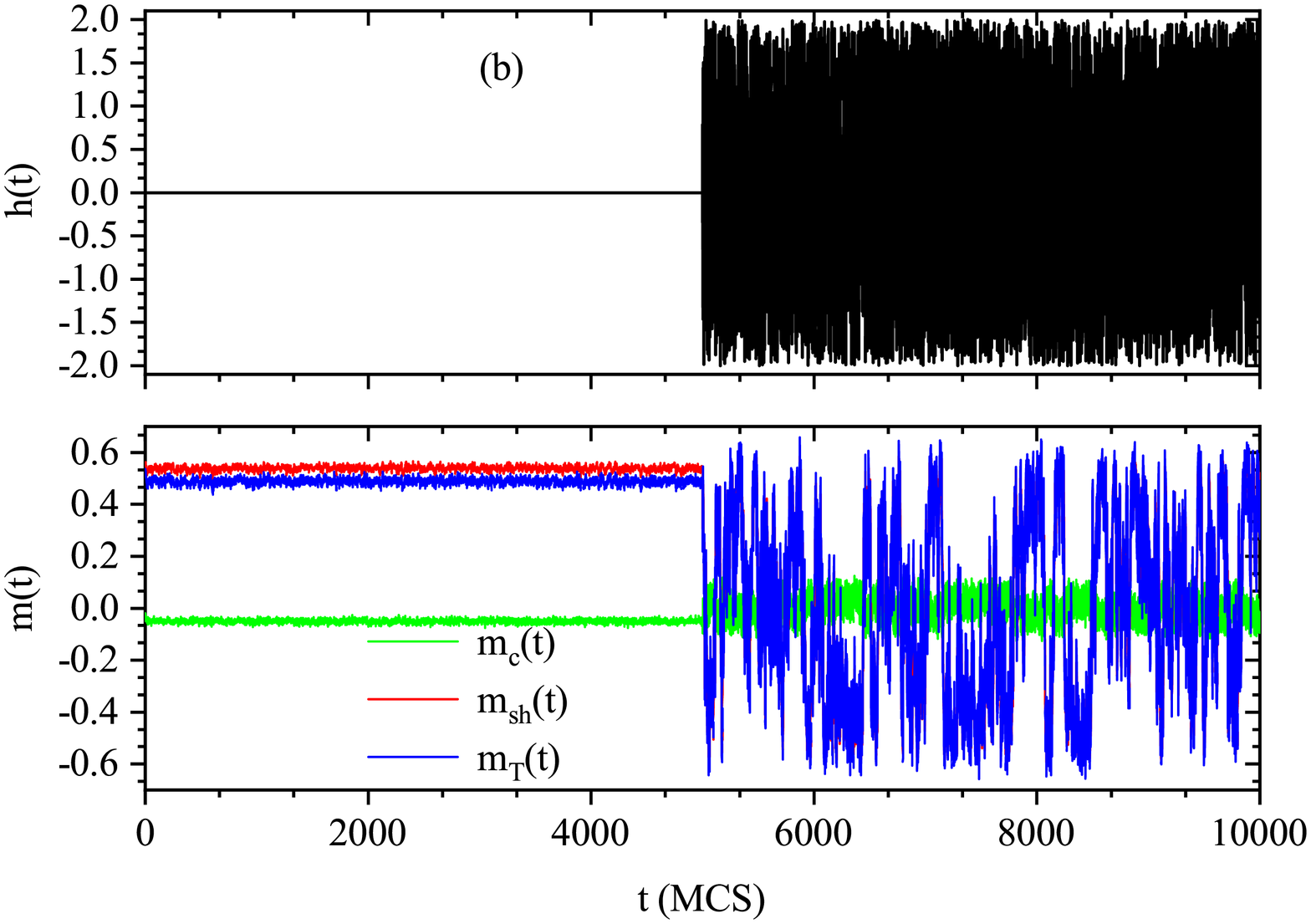}
\caption{\label{Fig1} (Color online) Monte Carlo results of the time variations of  randomly changing magnetic 
field and core, shell and total magnetizations. All results are obtained for a fixed temperature $k_{B}T/J_{sh}=2.0$ and 
the chosen values of the spin-spin couplings $J_c/J_{sh}=0.75$ and $J_{int}/J_{sh}=-1.5$}. Magnetic field sources, (a) $h/J_{sh}=1.0$ and (b) $h/J_{sh}=4.0$,  are opened at $t=5000$ MCs. This diagram demonstrates the dynamical symmetry breaking associated to the transition.
\end{figure}

\begin{figure}[htbp]
\includegraphics*[width=9.0 cm]{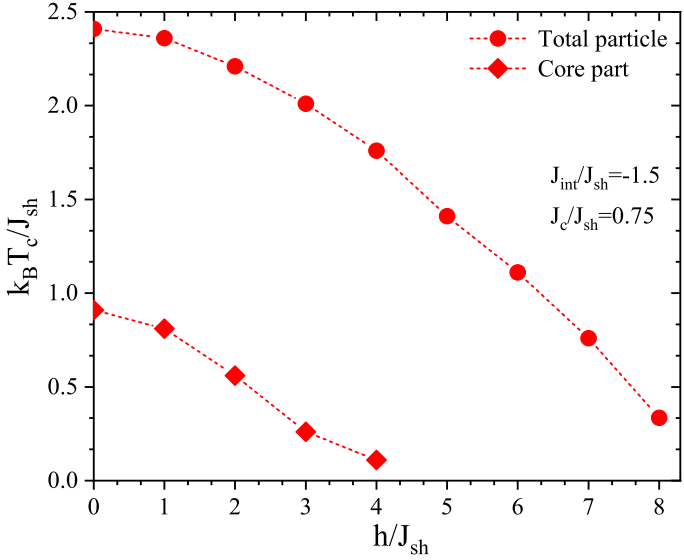}
\caption{\label{Fig2} (Color online) Dynamic phase boundary of core-shell nanocubic system 
in a $(h/J_{sh}-k_{B}T_{c}/J_{sh})$ plane for selected values of  $J_{c}/J_{sh}=0.75$ and $J_{int}/J_{sh}=-1.5$. The magnetic field
varies randomly between $-h/2J_{sh}$ and $h/2J_{sh}$. Transition points are obtained from the peak positions of the variances of the core, shell and total magnetizations and specific heat curve of the nanoparticle as a function of temperature. }
\end{figure}

\begin{figure}[htbp]
\includegraphics*[width=16.0 cm]{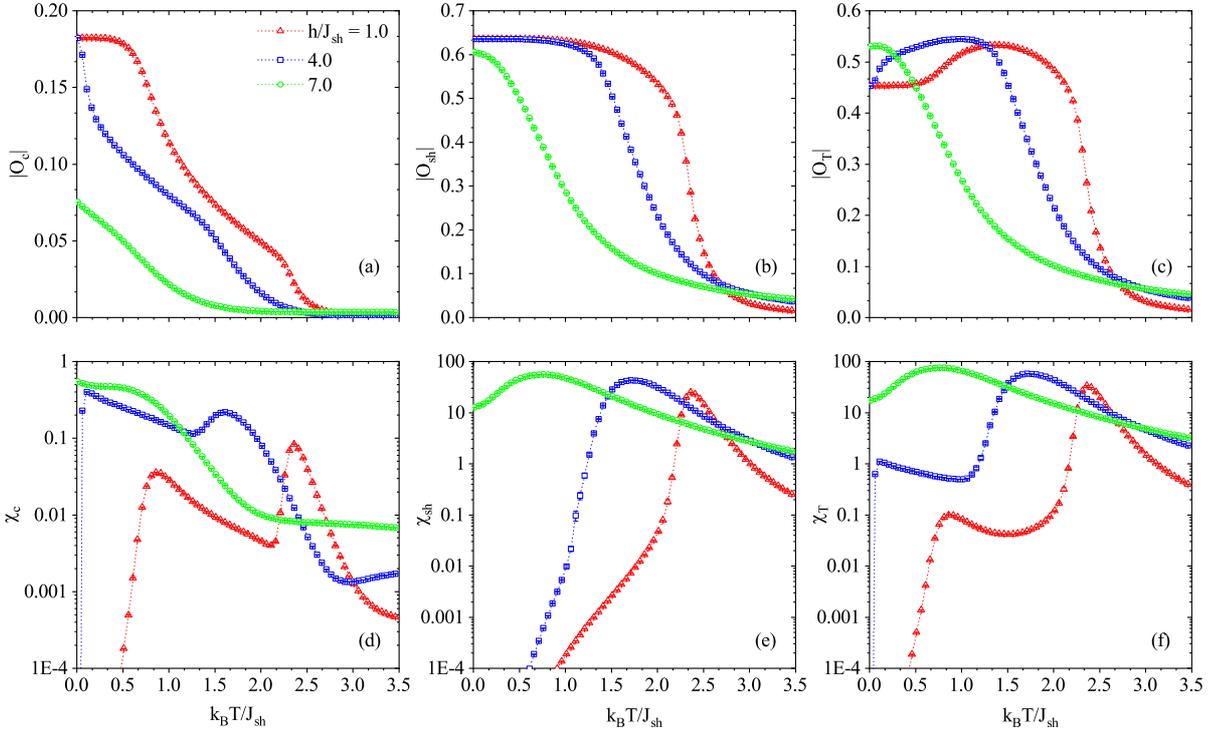}
\caption{\label{Fig3} (Color online) Effects of the randomly changing magnetic field width on the core (a), shell (b) and the total magnetizations (c), and their corresponding variances (d), (e) and (f), respectively.   Here, different symbols denote the varying magnetic values.}
\end{figure}

 As displayed in Figure \ref{Fig2}, in order to get a better understanding of the effect of randomly varying magnetic field width on the critical properties of the considered nanoparticle, we present the dynamic phase diagram plotted in a $(h/J_{sh}-k_{B}T_{C}/J_{sh})$ plane for a selected combination of system parameters such as $J_{int}/J_{sh}=-1.5$ and $J_{c}/J_{sh}=0.75$. Here, $T_{C}$ means the pseudo-critical temperature. Transition temperatures are extracted from the thermal variations of variances of the core, shell and total dynamic order parameters and internal energies. One of the outstanding 
results is that there is a multiple phase transition region including two branches 
in the system up to a particular magnetic field width. In this region, when the temperature increases starting from the relatively 
lower values, it has been found that the core magnetization first exhibits a phase transition. This corresponds to the first branch of the multiple transition line. Furthermore, if the temperature is increased  further, then the overall particle shows a transition between ordered and 
disordered phases for a fixed  value of $h/J_{sh}$. Such kind of a transition corresponds to the 
second branch of multiple transition line. 
Our Monte Carlo simulation findings show  that the multiple transitions observed here strongly depend on the chosen applied field width value.  As explicitly seen from the phase diagram, transition temperatures decrease 
when the $h/J_{sh}$ value gets bigger. This is because the energy contribution coming from the Zeeman term to the total 
energy increases with an increment in $h/J_{sh}$. Thereby, the phase transition 
lines gets shrink. We note that all of the transitions found here are second-order transition, i.e., 
there is no first-order phase transition in the system, indicating a tri-critical point where the first-order transition lines are met with the second-order transition lines. Similar kind of a multiple 
transition has also been reported in Ref. \cite{Acharyya2} where thermal and magnetic properties of a classical anisotropic Heisenberg model driven by a polarized magnetic field  are investigated by means of MC simulations.

\begin{figure}[htbp]
\includegraphics*[width=8.0 cm]{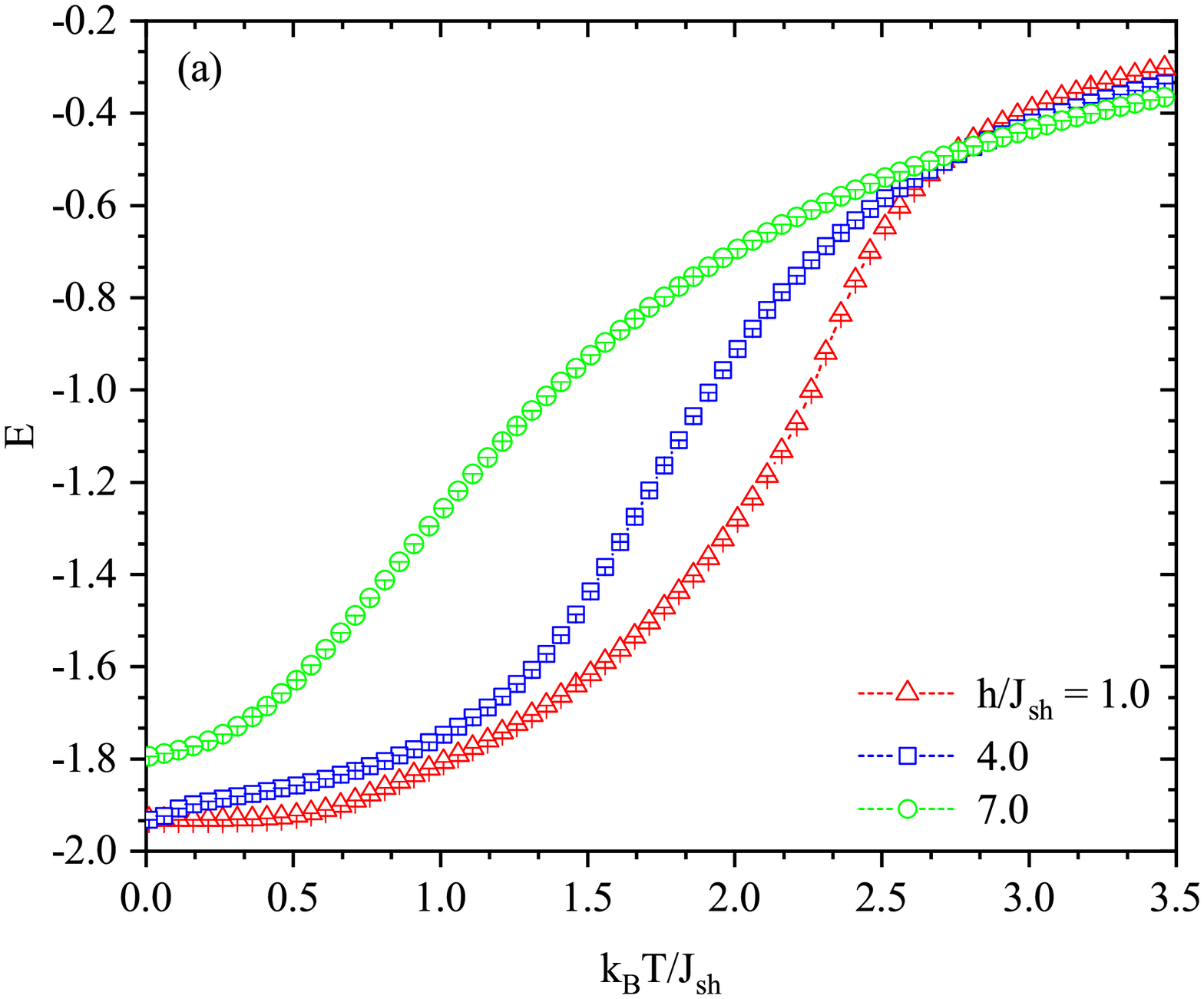}
\includegraphics*[width=8.0 cm]{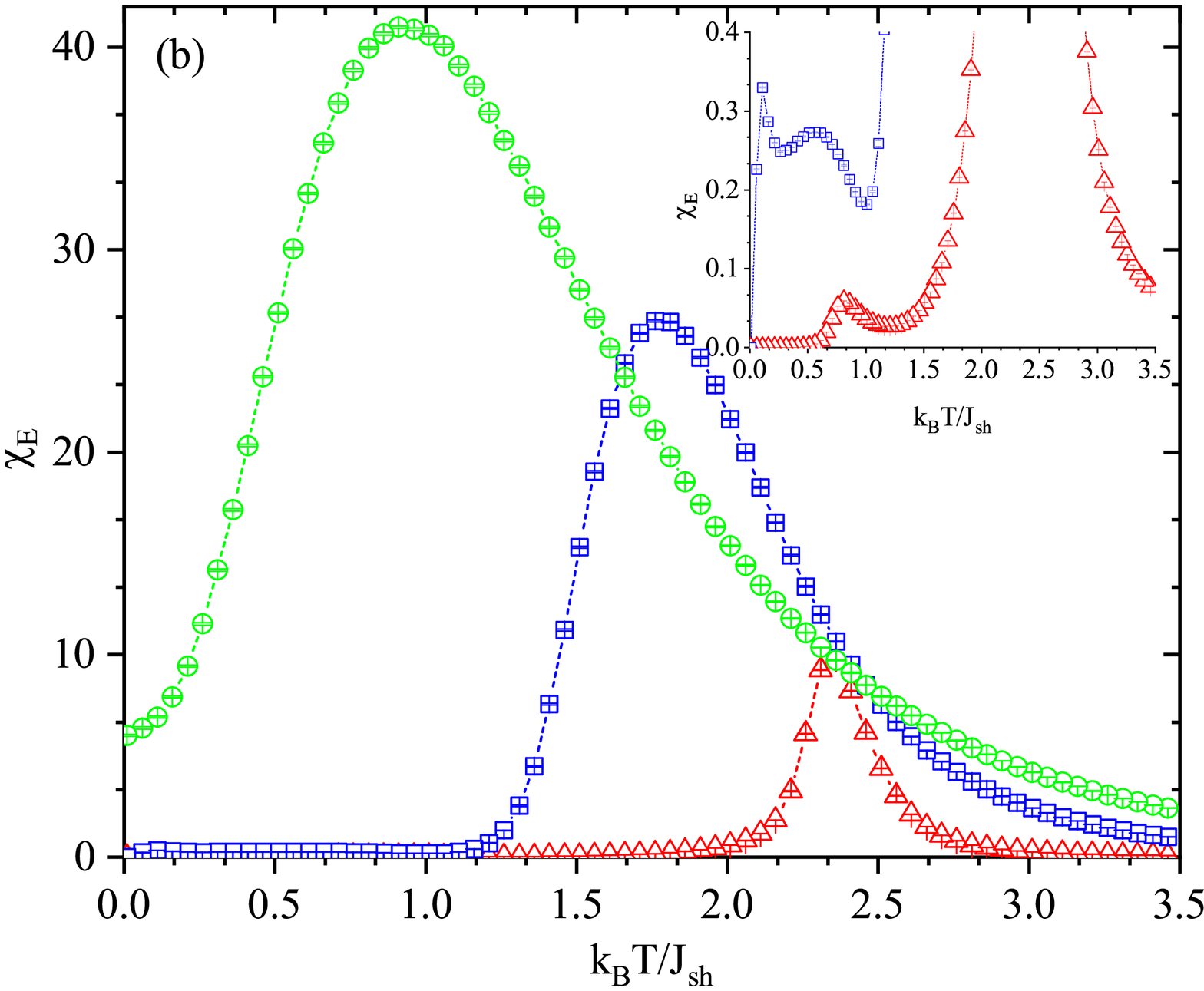}
\caption{\label{Fig4} (Color online) Thermal variations of the internal energy (a) and the corresponding variances (b). 
The curves are obtained for three different  values of the randomly changing magnetic field: $h/J_{sh}=1.0, 4.0$ and  $7.0$. Inset given  here represents the low temperature regions of the variances of the internal energies for two considered values of $h/J_{sh}$: $1.0$ and $4.0$.}
\end{figure}

In Figures \ref{Fig3}(a-c), we depict the effect of randomly changing magnetic field width on the core, shell and total dynamic order parameters as functions of the temperature, corresponding to the dynamic phase diagram given in Figure \ref{Fig2}. As an interesting observation, we can see from total magnetization curves plotted in Figure \ref{Fig3}(c), 
they exhibit a temperature induced maximum, which is strongly depend on the chosen applied field width. At this point we note that in the bulk ferrimagnetism of N\'{e}el \cite{Neel, Strecka}, it is possible to classify the magnetization profiles based on 
the total magnetization behaviors in certain categories.  According to this nomenclature, the system 
shows a P-type behavior for some selected values of $h/J_{sh}$ such as $1.0$ and $4.0$. It also means 
that the results given here are in common at various types of magnetic materials, which are not only 
dependent on their size. As the $h/J_{sh}$ gets bigger, for example $h/J_{sh}=7.0$, P-type behavior observed here tends to disappear, leading to a Q-type behavior. In Figures \ref{Fig3}(d-f), we present the thermal variations of the variances of core, shell and total dynamic order parameters for the same system parameters used in Figures \ref{Fig3}(a-c). As clearly seen from the figures that when the temperature reaches to the transition temperature, the related variances give rise to show a peak 
behavior, which is also a fingerprint of a continuous transition. Peak positions in $x$-axis are sensitively dependent on the studied external field width such that they tend to decrease with an increment in $h/J_{sh}$. Moreover, it is obvious from the Figure \ref{Fig3}(d) that the variance of core magnetization shows two peaks for some selected values of $h/J_{sh}$ such as $1.0$ and $4.0$. The first peak located at the relatively lower temperature region originates from the order-disorder transition of the core part of the nanoparticle whereas the second one is a result of the strong coupling between core and shell layers.

In Figure \ref{Fig4}, we give the thermal variations of the internal energy and corresponding variances for the same applied field width 
values used in Figure \ref{Fig3}. When the temperature begins to decrease starting relatively higher values, numerical values of the 
internal energies  are decreased for all values of $h/J_{sh}$. In addition to this general trend, as shown in Figure \ref{Fig4}(a), 
it is possible to say that a sudden decrement is observed when the temperature reaches to the relevant critical point, as in 
the case of equilibrium phase transitions \cite{Newman, Binder}. More specifically, figure \ref{Fig4}(b) displays the 
variances of the internal energies as functions of the temperatures. As in the case of the variances of the dynamic order parameters discussed above, when the temperature reaches to the critical point, they tend to show a clear peak behavior, indicating a second-order phase 
transition. Their positions on the $x$-axis are sensitively dependent on the considered applied field width values. In accordance with the previously observed results from the variances of the dynamic order parameters, the pseudo-critical temperatures start to shift to the 
lower regions with increasing $h/J_{sh}$ values.  In the inset, we also plot the low temperature behavior of the $\chi_{E}$ for two considered applied field width $h/J_{sh}$ such as $1.0$ and $4.0$. It is clear from the figure that, it shows a clear peak treatment supporting the transition of the nanoparticle belonging to only the core part, in the vicinity of the related transition point.  
\begin{figure}[htbp]
\includegraphics*[width=5.0 cm]{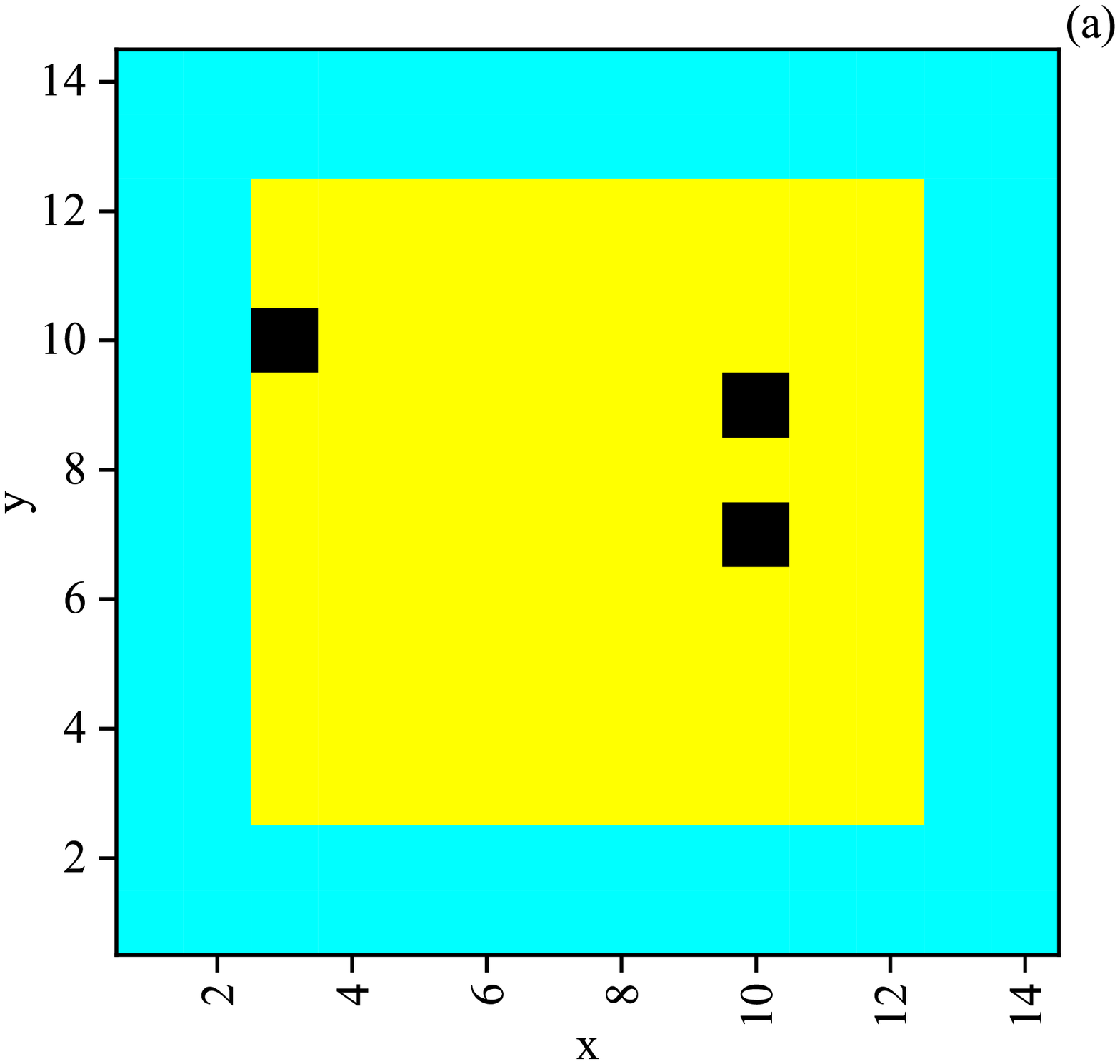}
\includegraphics*[width=5.0 cm]{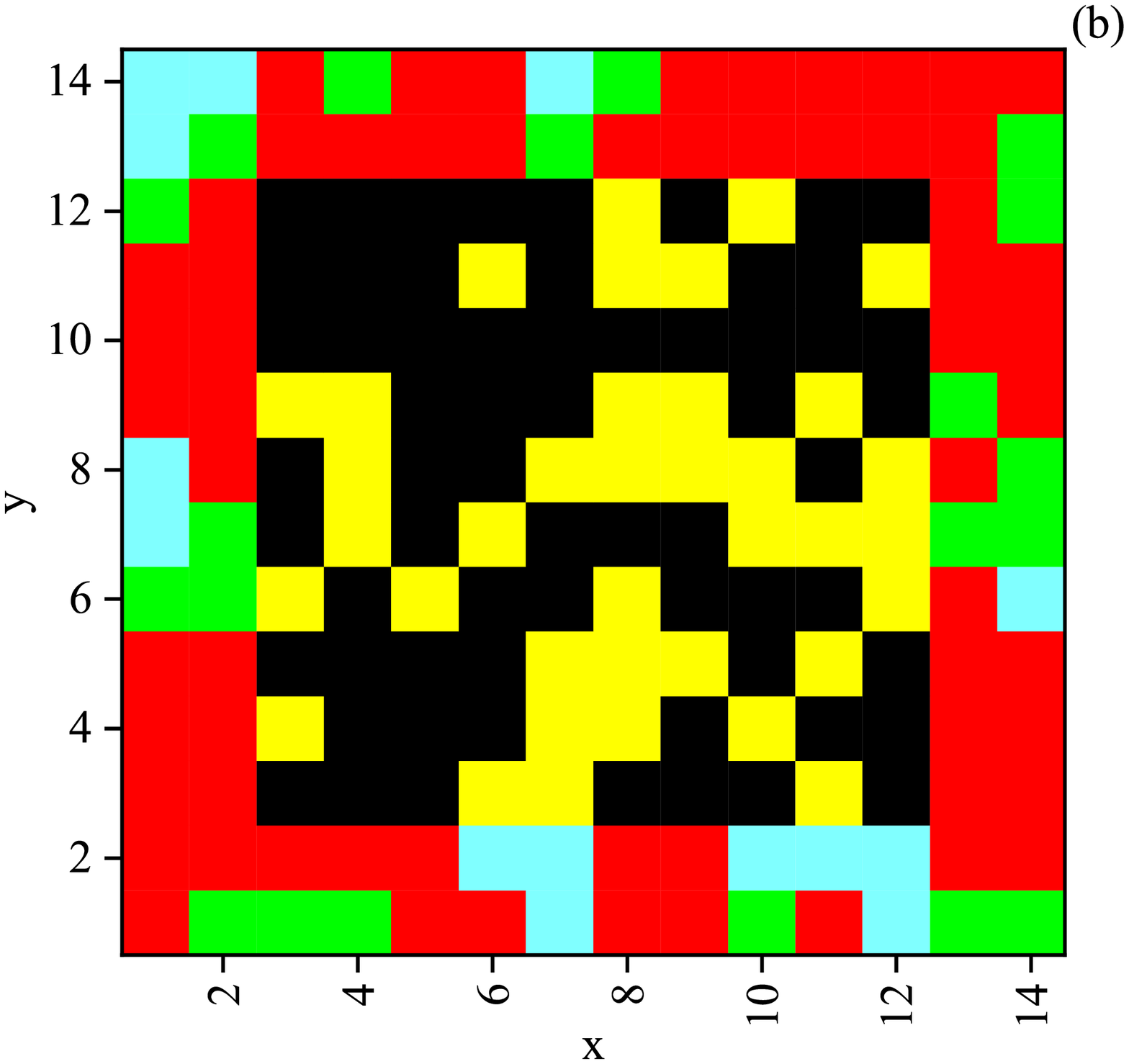}
\includegraphics*[width=5.5 cm]{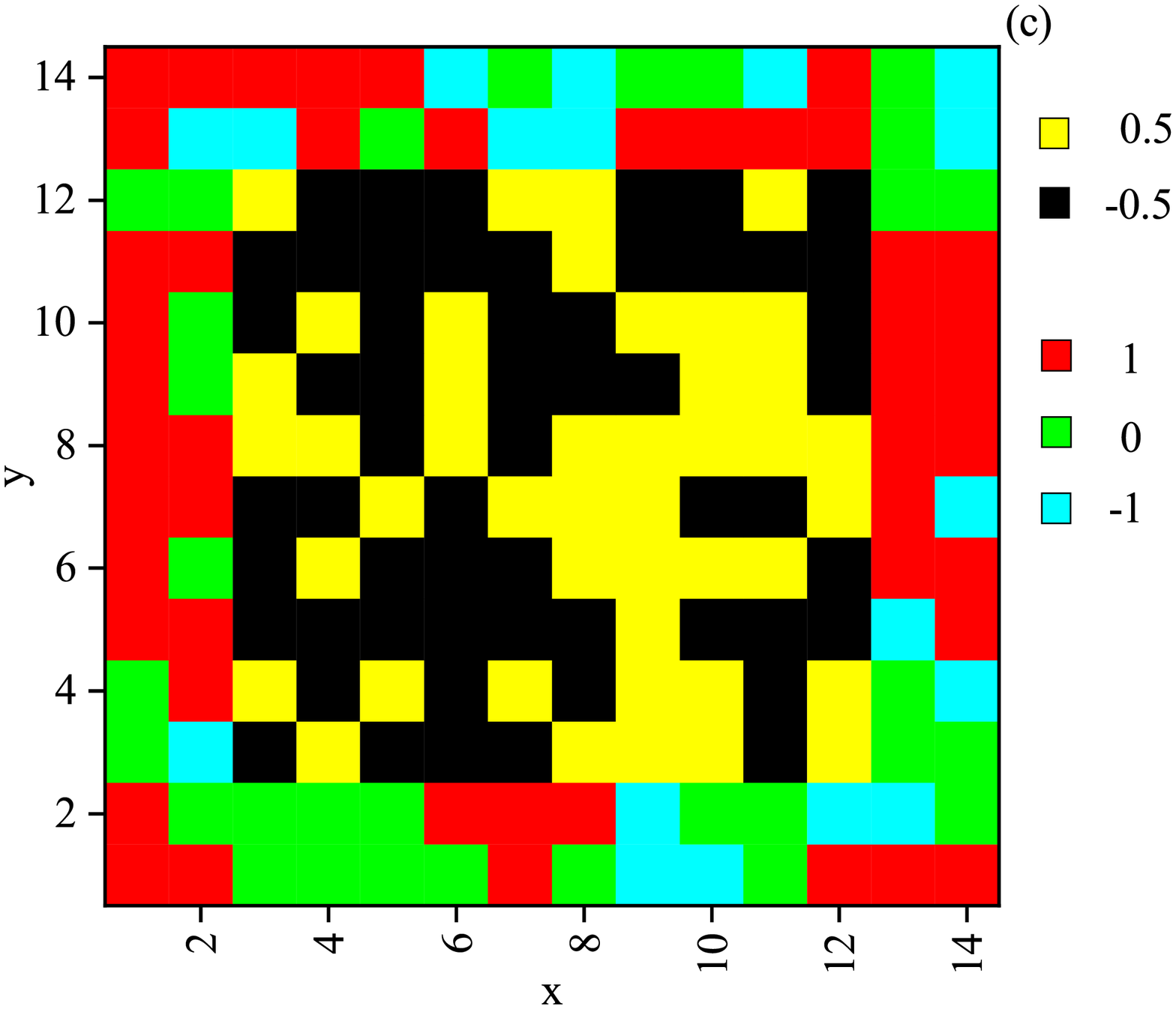}

\caption{\label{Fig5} (Color online) The spin configurations of the midplane cross-sections of the nanoparticle in the $x-y$ plane. They are taken for three different temperature values: (a) $k_{B}T/J_{sh}=0.5$ (ordered region), (b) $2.36$ (in the vicinity of transition point) and (c) 4.0 (disordered region) with $h/J_{sh}=1.0$. All snapshots are captured at $t= 75 \times 10^4$ MCs. Here, yellow, black, red, green and cyan colors correspond to the spin values of $0.5, -0.5, 1, 0$ and $-1$, respectively. }
\end{figure}

We represent the spin configurations of the midplane cross-sections of the nanoparticle in the $x-y$ plane in Figure
\ref{Fig5}. All snapshots are captured for the $h/J_{sh}=1.0$. By benefiting from the calculations mentioned above, the pseudo-critical is obtained as $k_BT_C/J_{sh}\approx2.36$ for $h/J_{sh}=1.0$. In order to show the influences of the temperature on the spin-snapshots, 
three different temperature values are taken into consideration, being smaller (a), equal (b) and larger than the critical 
temperature. As seen clearly from the figure \ref{Fig5}(a) that the spins in the core and shell parts of the system are opposite to each other due the existence of an antiferromagnetic exchange coupling betwen core and shell layers of the nanoparticle. However, when taking a look at the core and shell layers of the system one by one, it is seen that most of the spins in related layers are aligned parallel to each other, causing to an ordered phase. When the temperature reaches to the critical temperature, strong fluctuations take place in the nanoparticle, as depicted in Figure \ref{Fig5}(b). On the other hand, above $k_{B}T_c/J_{sh}$, the spins in the system try to follow randomly changing magnetic field. Therefore, they are aligned with respect to each other almost randomly [see Figure \ref{Fig5}(c)].

\begin{figure}[htbp]
\includegraphics*[width=9.0 cm]{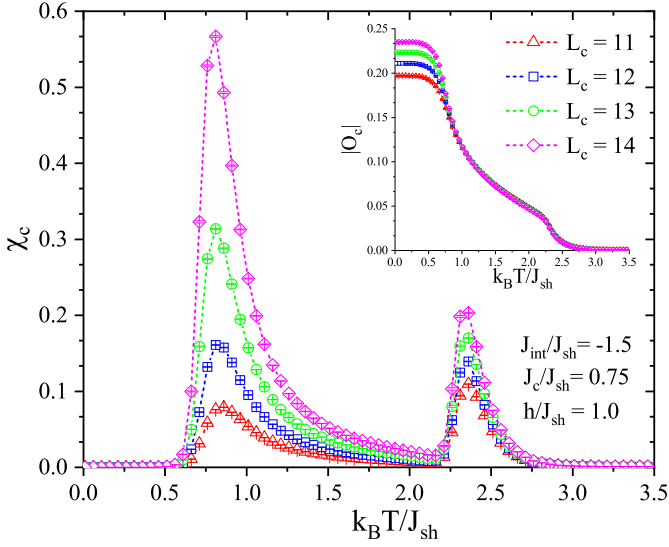}
\caption{\label{Fig6} (Color online) Thermal variations of the core susceptibilities 
for considered values of the system parameters such as $J_{int}/J_{sh}=-1.5, J_{c}/J_{sh}=0.75, h/J_{sh}=1.0$ and $L_{sh}=2$. 
The curves are plotted for four values of the core thickness: $L_c=11, 12, 13$ and $14$. The inset  
represents the temperature dependencies of the core magnetizations for the same parameter set used for the core susceptibilities curves. }
\end{figure}

 Figure \ref{Fig6} displays the influences of core thickness on the core 
susceptibilities (main panel) and the core magnetizations (inset) for considered values of the system 
parameters such as $J_{int}/J_{sh}=-1.5, J_{c}/J_{sh}=0.75, h/J_{sh}=1.0$ and $L_{sh}=2$. The curves are 
obtained for four values of $L_c=11, 12, 13$ and $14$. It can be deduced from the figure 
that the multiple transition behavior is not affected by the variation of the core thickness. 
When $L_c$ gets bigger, the height of the core susceptibilities even prominently increases, leading to a strong 
ferromagnetic character in the core part of the nanoparticle. The behavior mentioned here is also seen in the curves of 
the core magnetization as a function of the temperature. Although all curves qualitatively are similar to 
each other, the saturation values sensitively depend on the chosen value of $L_{c}$. Moreover, in order to 
see whether the multiple phase transition character obtained here depends on the sign of the interface coupling 
parameter or does not, we give again the core susceptibilities (main panel) and core 
magnetizations (inset) versus temperature curves, as displayed in Figure \ref{Fig7}. 
We changed only the sign of the $J_{int}/J_{sh}$ parameter without changing its magnitude for the same system 
parameters used for Figure \ref{Fig6} with $L_c=10$.  It is clear from the figure that thermal variations of 
the core susceptibilities show a double-peak behavior, and the phase transition temperatures corresponding to 
the core part and overall nanoparticle are almost the same for both  values of  $J_{int}/J_{sh}=1.5$ and $-1.5$. The 
inset displays the thermal variations of the core magnetizations for the same parameter set. We note that there 
is a clear agreement between these two curves obtained for the values of  $J_{int}/J_{sh}=1.5$ and $-1.5$. We also checked the 
remaining quantities such as the shell and the total order parameters, corresponding susceptibilities, the internal 
energies and the specific heat curves as functions of the temperature, which are not given here, 
and  almost the same results have been obtained for both values of $J_{int}/J_{sh}$. In view of all these 
observations, we can say that a variation in the sign of the interface coupling, $J_{int}/J_{sh}\rightarrow - J_{int}/J_{sh}$, 
does not  significantly affect the multiple transition behavior reported here. We note that such kind of studies 
considering the effects of the sign of the interface coupling parameter on the magnetic 
properties of the different kinds of core/shell nanoparticle systems have been elucidated in the 
references \cite{Iglesias, Yuksel2}.

\begin{figure}[htbp]
\includegraphics*[width=9.0 cm]{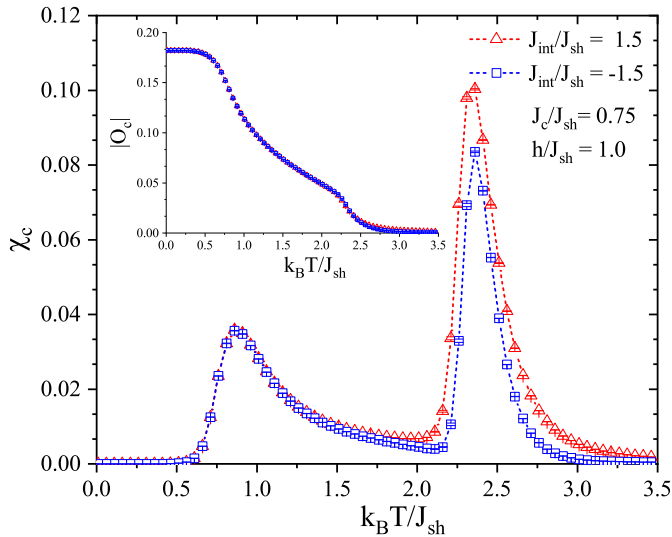}
\caption{\label{Fig7} (Color online) Monte Carlo simulation results of the thermal variations of core 
susceptibilities of the nanoparticle for varying values of $J_{int}/J_{sh}$. 
All results are obtained for some selected values of the system parameters such as 
$h/J_{sh}=1.0, J_c/J_{sh}=0.75, L_{sh}=2$ and $L_c = 10$. The corresponding core parts of 
the nanoparticle for the same parameter set are displayed in the inset.}
\end{figure}

 As a final investigation, in order to show the effects of the varying system size of the nanoparticle on its thermal and 
magnetic properties, we depict the dynamic phase boundary in a $(L_{sh}-k_BT_{c}/J_{sh})$ plane in Figure \ref{Fig8}(a)  for 
considered values of the 
system parameters: $J_{int}/J_{sh}=-1.5$, $J_{c}/J_{sh}=0.75$ and $h/J_{sh}=1.0$. We followed the same way mentioned above 
to extract the phase transition points. It is worthwhile to mention that there is still a multiple phase transition behavior 
including two branches in the system for all selected values of $L_{sh}$. More specifically, the phase 
transition point regarding the overall nanoparticle tends to increase with increasing $L_{sh}$ starting from $L_{sh}=2$, 
and it saturates to a certain value with further increment in $L_{sh}$. It means that after a certain 
value of $L_{sh}$ the boundary conditions begin to lose its effect on the system, and the systems starts to behave 
like a bulk system. It should be noted that varying shell thickness values do not affect the first branch of the 
phase diagram where the core magnetization shows a phase transition, in accordance with the expectations. 
In other words, the critical temperature for core part of the nanoparticle is insensitive to the shell thickness of the system. As seen 
in Figures \ref{Fig8}(b) and (c), varying shell thickness values affect clearly the core, shell and total 
magnetizations of the system. The saturation value of the shell magnetization increases when $L_{sh}$ gets bigger. Another important findings 
is that the total magnetization as a function of the temperature 
starts to evolve from the P-type to Q-type behavior with increasing $L_{sh}$ \cite{Strecka}.

\begin{figure}[htbp]
\includegraphics*[width=18.0 cm]{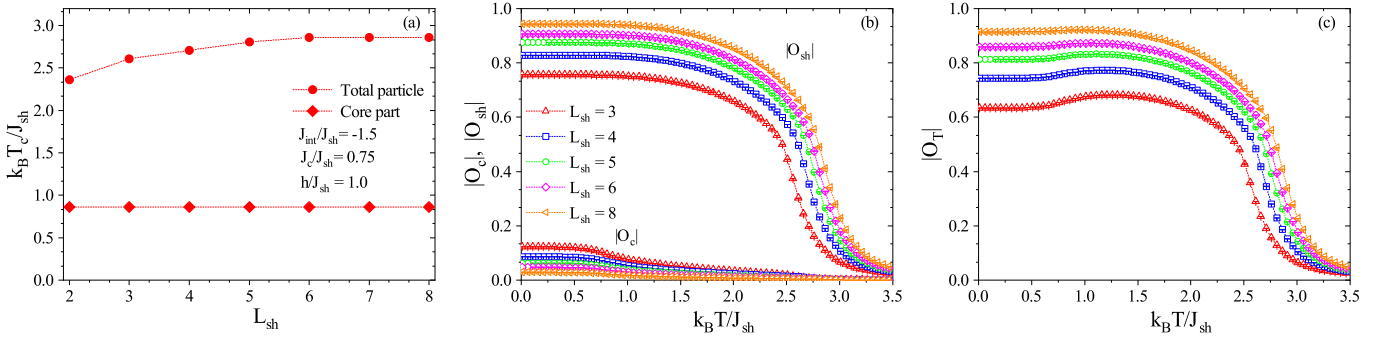}
\caption{\label{Fig8} (Color online) (a) Shell thickness dependencies of the dynamic critical points of a cubic core/shell
nanoparticle for considered values of the system parameters such as $J_{int}/J_{sh}=-1.5, J_{c}/J_{sh}=0.75$ and $h/J_{sh}=1.0$. 
The effects of the shell thickness on the thermal variations of core and shell (b) and total (c) magnetizations at varying values 
of the shell thickness $L_{sh}=3, 4, 5, 6$ and $8$. }
\end{figure}

\section{Concluding Remarks}\label{Conclusion}

 The equilibrium behaviours of the core-shell nanoparticles have been investigated widely
and well understood.
However, the nonequilibrium responses of the core-shell nanoparticles have not yet been
investigated. Here we have studied the nonequilibrium responses of Ising core-shell
nanoparticles driven by randomly varying (in time but uniform over the space) magnetic
field. The time dependent magnetic field keeps the core-shell nanoparticle
far from equilibrium. Hence, the responses of the system is truely of nonequilibrium 
type. The ferromagnetic Ising (spin-1/2) core is covered by ferromagnetc Ising (spin-1)
shell. However, the interfacial interaction between core and shell is being considered
as antiferromagnetic. We have calculated the time averaged core magnetization,
shell magnetization and the total magnetization and their variances (susceptibility in
a sense) and studied those as functions of the temperature with field as parameter.
We have observed, for relatively higher value of the temperature and the width of
the randomly varying magnetic field, the system shows a dynamical symmetric behaviour.
In this case, the instantaneus magnetizations are found to fluctuate symetrically
around zero value. On the other hand, for relatively lower values of the temperature
and the field width a dynamically symmetry broken phase is observed. Where the
instantaneus magnetizations fluctuates asymmetrically around zero value. These are
demonstrated in figure-\ref{Fig1}. Obviously,
a nonequilibrium phase transition is observed in association with a dynamical symmetry
breaking although the concept of symmetry breaking was originally made for the system in thermodynamic limit. 
In the above mentioned dynamically symmetric phase, the time averaged magnetizations 
(serving as the order parameters in the present study) vanishes, indicating a nonequilibrium 
disordered phase. Whereas, in the dynamically asymmetric phase, the time averaged magnetizations, 
acquires nonzero values, indicating the 
nonequilibrium ordered phase. The transition temperatures were estimated from the
value of the temperatures where the variances of those magnetizations get peaked
(believed to be diverging in the thermodynamic limit). It was observed, that
the variance of total magnetization gets peaked at higher temperature and that of
core magnetization gets peaked at some lower temperature. This is a signature of
{\it nonequilibrium multiple phase transition}
 observed in the driven core-shell nanoparticle.

Collecting all these nonequilibrium transition temperatures the comprehensive
nonequilibrium phase diagarm is drawn in the plane consists with the axes of
the temperature and the width of the randomly varying magnetic field.

It would be quite interesting to study such behaviours in the case of Ising
core-shell nanoparticles where the values, of spins in the core and that in the shell,
differes largely. That will affect certainly on the difference in the region 
bounded by core-transition line and the shell-transition line in the comprehensive
phase diagram. What will be the effects of dilution here ? If the core and shell
are diluted by nonmagnetic impurities (with different concentrations in core and
shell), how does it affect the phase boundaries of the multiple phase diagram.
Last but not the least, how sensitive the nature of the distribution (i.e.,uniform,
bimodal, normal) of the randomly varying magnetic fields, would also be an interesting matter to be studied. We have plans
to study these and the results will be reported elsewhere.

\section*{Acknowledgements}
The numerical calculations reported in this paper were performed at TUBITAK ULAKBIM High Performance and Grid Computing Center (TR-Grid e-Infrastructure). MA acknowledges
FRPDF research grant provided by Presidency University.

\end{document}